\documentclass[12pt]{article}
\pdfoutput =1
\usepackage{float} 
\usepackage{amsmath}
\usepackage{amsfonts}   
\usepackage{amssymb}

\begin{document}
\date{}
\title{{\bf{\Large Hydrodynamics from scalar black branes}}}
\author{
 {\bf {\normalsize Dibakar Roychowdhury}$
$\thanks{E-mail:  dibakarphys@gmail.com, dibakar@cts.iisc.ernet.in}}\\
 {\normalsize Centre for High Energy Physics, Indian Institute of Science, }
\\{\normalsize C.V. Raman Avenue, Bangalore 560012, Karnataka, India}
}

\maketitle
\begin{abstract}
In this paper, using the Gauge/gravity duality techniques, we explore the hydrodynamic regime of a very special class of strongly coupled QFTs that come up with an emerging UV length scale in the presence of a negative hyperscaling violating exponent. The dual gravitational counterpart for these QFTs consists of scalar dressed black brane solutions of exactly integrable Einstein-scalar gravity model with Domain Wall (DW) asymptotics. In the first part of our analysis we compute the $ R $-charge diffusion for the boundary theory and find that (unlike the case for the pure $ AdS_{4} $ black branes) it scales quite non trivially with the temperature. In the second part of our analysis, we compute the $ \eta/s $ ratio both in the non extremal as well as in the extremal limit of these special class of gauge theories and it turns out to be equal to $ 1/4\pi $ in both the cases. These results therefore suggest that the quantum critical systems in the presence of (negative) hyperscaling violation at UV, might fall under a separate universality class as compared to those conventional quantum critical systems with the usual $ AdS_4 $ duals.
 \end{abstract}

\section{Overview and Motivation}
For the past several years, the holographic description of AdS black branes in the presence of non trivial scalar profile has drawn renewed attention due to its rich phenomenological contents in the context of  AdS/CFT  duality where the scalar hair in the bulk plays the role of the (scalar) condensation operator in the dual field theory \cite{Charmousis:2010zz}-\cite{Iizuka:2011hg}. It has been observed that due to the presence of the (non)minimal coupling between the scalar field and the $ U(1) $ gauge field(s), the near horizon (IR) physics comes up with an emerging length scale close to the the extremal limit of the brane. In other words, Quantum Field Theories (QFTs) dual to these gravitational theories exhibit the so called \textit{scale covariance} in the IR which decouples as one RG flows towards the UV fixed point of the theory \cite{Cadoni:2011nq}-\cite{Bhattacharya:2012zu}. In the IR, such QFTs are essentially described in terms of two parameters, namely the dynamic critical exponent ($ z $) and the \textit{hyperscaling} violating parameter $ \theta (> 0)$. Such hyperscaling violating  QFTs play an extremely important role in the holographic understanding of certain condensed matter phenomena and beyond that, for example the holographic description of the so called Fermi surfaces \cite{Huijse:2011ef}-\cite{Li:2012uua} as well as the area law violation in the context of holographic entanglement entropy \cite{Shaghoulian:2011aa}-\cite{Kulaxizi:2012gy}.

One of the major reasons behind the existence of such hairy configurations rests on the fact that the interaction potential $ V(\phi) $ (in the presence of the mild Tachyonic excitations which thereby makes the usual charged black brane configurations unstable below certain critical value of the temperature) exhibits a negative local maxima for the vanishing of the scalar field ($ \phi $). As a continuation of our discussion on the hairy configurations, it is noteworthy to mention that recently there has been some investigations in order to explore whether one could construct hairy configurations in the presence of positive mass squared ($ m^{2}>0 $) excitations by considering \textit{non} AdS asymptotic boundary conditions. These analysis suggest that one can in fact construct scalar dressed black brane configurations in the context of exactly integrable Einstein-Scalar gravity with positive mass squared excitations and these are termed as Scalar Black Branes (SBB) in the literature \cite{Cadoni:2011yj}-\cite{Cadoni:2012ea}. 

The most striking feature about these SBBs is the existence of the interpolating soliton between a scale covariant UV ($ z=1 $,~$ \theta <0 $) and a scale invariant IR fixed point ($ z=1 $, $ \theta =0 $) in the extremal limit of the brane \cite{Cadoni:2011yj}-\cite{Cadoni:2012ea}. From the free energy ($ F=M-TS $) computations it is in fact easy to note that the SBB configuration is stable/ thermodynamically preferred only above certain critical temperature ($ T>T_c $). Otherwise, at low temperatures it is the Schwarzschild AdS (SAdS) solution (with the vanishing of the scalar field ($ \phi =0 $)) that turns out to be thermodynamically most preferred configuration in the bulk \cite{Cadoni:2012uf}.  

This is precisely the place where we end up with certain specific motivation behind our present analysis. Before we actually come to that, let us first note that the QFTs dual to SBBs exhibit a strikingly different feature, namely they come up with certain emerging length scale at UV (which essentially decouples in the IR) in the presence of a negative hyperscaling violating exponent ($ \theta <0$) \cite{Cadoni:2012uf}. This feature is indeed opposite to that of the earlier holographic scenarios \cite{Dong:2012se},\cite{Kim:2012nb},\cite{Huijse:2011ef}. This is also quite unusual from the point of view of the QFTs encountered in the usual condensed matter systems in the sense that there one generally expects the hyperscaling violation to occur in the IR and not in the UV scale of the theory \cite{Fisher:1986zz}. As a matter of fact, the dramatic change in the boundary behaviour of the scalar field (for $ T>T_c $) corresponds to turning on different operators for the boundary field theory which in turn suggests that QFTs dual to SBBs are eventually described in terms of a different Lagrangian compared to those dual to usual SAdS for $ T<T_c $ \cite{Cadoni:2012ea}.
Another important issue that should be noted at this stage is the fact that due to the presence of the negative hyperscaling violating exponent ($ \theta $) these QFTs cannot be fitted into the holographic framework of the so called condensed matter systems in ($ 2+1 $) dimensions since the hyperscaling violating factor essentially changes the dimensionality ($ d $) of the system to $ d-\theta $ which therefore in the present scenario seems to increase the effective dimensionality of the system\footnote{Holographic models with negative values of $ \theta $ have been constructed earlier in the context of $ Dp $ branes in \cite{Dong:2012se}. The rising of the effective dimensionality in the boundary field theory seems to indicate the fact that the dual gravitational counterpart would prefer to live in more than ($ 3+1 $) dimensions \cite{Cadoni:2012uf}. Therefore from the point of view of the boundary field theory these SBBs are really interesting objects and worth of further investigations.}. 

We would like to emphasise these points a bit further. We would like to ask the following question: How does the two point correlation between two operators namely, $  \langle \mathcal{O}_{i}(\textbf{x})  \mathcal{O}_{j}(0)\rangle $ behaves in the low frequency (\textit{hydrodynamic}) limit of these special class of strongly coupled QFTs at UV. In other words, the purpose of the present article is to use the usual frame work of the Gauge/gravity duality in order to explore the low frequency behaviour of the two point correlators for a very special class of strongly coupled hyperscaling violating QFTs both in the extremal as well as in the non extremal limit\footnote{In \cite{Cadoni:2012ea}, the authors have studied the short distance and/or the high frequency behaviour of the two point correlation function between the scalar operators for these special class of QFTs ( with $ \theta <0$) and observed the power law behaviour in the two point function. This observation leaves the ground open for further investigations on the behaviour of these correlation functions in the low frequency regime.}.

Let us illustrate this point a bit further. The goal of the present article is to study the physics of linear response for a special class of strongly coupled hyperscaling violating QFTs at sufficiently long length scales and to check the universality between its various response parameters, namely the charge conductivity ($ \sigma_{DC} $) and the shear viscosity ($ \eta $). 

In order to check these universalities, we essentially compute the transport coefficients and/or the response parameters for the boundary hydrodynamics using the so called Kubo's formula,
\begin{eqnarray}
\chi_{\Re} =-\lim_{\omega\rightarrow 0}\frac{1}{\omega}\mathcal{G}^{R}(\omega,\textbf{k}=0)
\end{eqnarray}
where $ \mathcal{G}^{R}(\omega,\textbf{k}) $ is the retarded Green's function that relates the boundary operator $ \mathcal{O} $ to its classical source $ \varphi_B $.
In our analysis, we systematically address the above universality issue(s) in two steps. In the first part of our analysis, we compute the $ R $- charge diffusion \cite{Kovtun:2003wp}-\cite{Policastro:2002se} for the boundary hydrodynamics where we turn on $ U(1) $ fluctuations over the fixed back ground of the SBBs and study the behaviour of the $ \langle J_{\mu}(\textbf{x})\ J_{\nu}(0) \rangle $ correlators in the low frequency limit which enable us to compute the $ \sigma_{DC}/\chi $ ratio for these special class of strongly coupled QFTs near its UV scale. In the remaining part of our analysis, 
 we turn on massless spin 2 (graviton) excitations in the bulk and compute the correlators of the type $ \langle T_{\mu\nu}(\textbf{x})\ T_{\varrho\sigma}(0) \rangle $  in the low frequency limit. From our analysis we observe that (like in the case for the pure $ AdS_{4} $ black branes) the low frequency (near horizon) graviton fluctuations in the bulk capture the physics of linear response for the boundary hydrodynamics and the corresponding universality of the $ \eta/s $ ratio \cite{Policastro:2001yc}-\cite{Iqbal:2008by} is guaranteed even in the presence of the negative hyperscaling violating effects near the UV scale of the theory. In summary, we find that the universality of the $ \eta/s $ ratio is still maintained both in the extremal as well as in the non extremal limit, while on the other hand, such universality relations do not seem to hold for the $ R $- charge diffusion. In fact we note that (unlike the case for the pure $ AdS_{4} $ black branes) the latter quantity exhibits a non trivial scaling relation with the temperature. These observations therefore enforce us to conclude that the (hyperscaling violating) gauge theories dual to SBBs might fall under a separate universality class as compared to those QFTs dual to usual $ AdS_4 $ black branes. 

Before we conclude our introductory section, it is customary to mention some the notable facts about our analysis. Let us first focus in particular in the \textit{extremal} limit. Note that taking the hydrodynamic limit of near extremal $ AdS_4 $ black branes has been always a bit tricky since the $ \omega\rightarrow 0 $ and the $ T\rightarrow 0 $ limits do not commute in general. On top of it, due to the presence of the double pole structure in the near horizon limit, the general arguments concerning the universality of the $ \eta/s $ ratio in the non extremal limit should not necessarily hold near the extremal point \cite{Edalati:2009bi}-\cite{Paulos:2009yk}. Considering all these facts the present scenario for our analysis turns out to be more delicate in the sense that the near horizon geometry of the \textit{extremal} soliton (dressed with scalar hair) is radically different (it does not contain the usual $ AdS_2 $ factor) from that of the extremal $ AdS_4 $ black branes \cite{Edalati:2009bi}-\cite{Paulos:2009yk}. Therefore from the bulk point of view it appears that the usual arguments that go in favour of extremal $ AdS_4 $ black branes should not directly apply here. However, we figure out that due to the presence of the double pole structure near the horizon of the extremal SBBs one recovers the usual lower bound for $ \eta/s $ \cite{Edalati:2009bi}-\cite{Paulos:2009yk}. From the boundary point of view this simply indicates the fact that these strange quantum critical fluids with $ \theta<0 $, exhibit identical shear as those of the usual quantum critical systems with $ \theta \geq 0 $ \cite{Dong:2012se},\cite{Kim:2012nb},\cite{Huijse:2011ef}. Therefore from the point of view of the boundary hydrodynamics the analysis near the extremal limit appears to be more non trivial as well as worthy of further investigations.

The organisation of the paper is the following. In Section 2, we start with a brief introduction to the dual gravitational description in the bulk and discuss its supergavity realizations. In Section 3, following the original prescription of \cite{Kovtun:2003wp}-\cite{Policastro:2002se}, we compute the $ R $- charge diffusion constant for the dual field theory in its UV limit. In Section 4, considering the non extremal SBBs, we check the universality of $ \eta/s $ ratio for these special class of gauge theorise at strong coupling regime. 
In Section 5, we perform the $ \eta/s $ computation in the extremal limit. Finally, we conclude in Section 6.

\section{Review of the gravity dual}
We start our analysis with a brief introduction to the fundamentals of the gravity set up in the bulk (in particular in the \textit{non extremal} limit) which essentially consists of Einstein's gravity minimally coupled with the scalar field ($ \phi $) in the presence of a self interaction potential ($ V(\phi) $) namely \cite{Cadoni:2012uf}, 
\begin{eqnarray}
S &=& \int d^{4}x\sqrt{-g} (\mathcal{R} -2 (\partial \phi)^{2}-V(\phi))\nonumber\\
V(\phi)&=& - \frac{6}{\gamma L^{2}}\left( \exp{2\sqrt{3}\beta \phi}-\beta^{2}\exp{\frac{2\sqrt{3}\phi}{\beta}}\right);~~\gamma =1-\beta^{2}\label{E1}
\end{eqnarray}
where $ L $ is the fundamental length scale of the $ AdS_4 $ and $ \beta $ is a real parameter such that for the present analysis of this paper $ |\beta| \ll 1 $. Before we proceed further, a few crucial points are to be noted at this stage. First of all, the above action in (\ref{E1}) essentially gives rise to two classes of static black brane configurations, one is the Schwarzschild black brane (SAdS) configuration (for $ T<T_c $) which corresponds to the vanishing of the scalar field along with some local minima in the potential, $ V=-6/L^{2} $ and the other corresponds to Scalar Black Brane (SBB) solution in the presence of the non trivial scalar profile\footnote{For details see \cite{Cadoni:2012uf}. },
\begin{eqnarray}
ds^{2}&=&\Delta^{\frac{2\beta^{2}}{3 \gamma}}\left[ \left(\frac{r}{r_0} \right)^{\frac{2}{1+3\beta^{2}}}(-\Gamma dt^{2}+dx_{i}^{2}) +\frac{E \Delta^{\frac{4\beta^{2}}{3 \gamma}}}{\Gamma}\left(\frac{r}{r_0} \right)^{\frac{-2}{1+3\beta^{2}}}dr^{2}\right]\nonumber\\
\phi &=& \frac{\beta}{\sqrt{3}\gamma} \log \left[ \frac{A}{\Delta}\left(\frac{r}{r_0} \right)^{\frac{-3\gamma}{1+3\beta^{2}}}\right].\label{E2} 
\end{eqnarray}

The SBB solution (\ref{E2}) could be found to possesses a number of interesting features. For example, in the extremal limit, it is in fact possible to show that for certain choice of parameters in the theory, the SBB solution depicted above in (\ref{E2}) could be thought of as a scalar soliton interpolating between an IR fixed point (that could be thought of as an $ AdS_{4} $ with the AdS length scale $ L $) and the \textit{scale covariant} Domain Wall (DW) kind of asymptotics at UV. In other words, in the UV we encounter certain hyperscaling violation due to some emerging UV length scale that essentially decouples in the IR limit. At this stage it is noteworthy to mention that the IR ($ r\rightarrow 0 $) limit of the soliton (\ref{E2}) essentially corresponds to the local minima of the potential namely, $ V=-6/L^{2} $. On the other hand, the UV ($ r\rightarrow \infty $) limit stands for the zero of the potential. At this stage it is noteworthy to mention that one can in fact express (\ref{E2}) in a more compact form using certain dimensionless coordinates ($ t,r $) namely \cite{Cadoni:2012uf},
\begin{eqnarray}
ds^{2}&=& \frac{\gamma^{2}L^{2}}{(1+3\beta^{2})^{2}}\left[- \Delta^{\frac{2\beta^{2}}{3\gamma}}\Xi(r)dt^{2} + \Delta^{\frac{2\beta^{2}}{\gamma}}\Xi^{-1}(r)dr^{2}\right] + \Delta^{\frac{2\beta^{2}}{3\gamma}}r^{\frac{2}{1+3 \beta^{2}}}(dx^{2}+dy^{2})\nonumber\\
\phi &=& \frac{1}{2}\log (\Delta^{\frac{-2\beta}{\sqrt{3} \gamma}}r^{\frac{-2\sqrt{3}\beta}{1+3\beta^{2}}});~~\Xi =r^{\frac{2}{1+3 \beta^{2}}}\left( 1- \frac{\Theta_{1}}{r^{\alpha}}\right);~~\Delta(r)=1+\frac{\Theta_{2}}{r^{\alpha}}\nonumber\\
\alpha &=&\frac{3\gamma}{1+3\beta^{2}}\label{E3}
\end{eqnarray}
where $ \Theta_{1,2} $ are two dimensionless parameters\footnote{One important assumption of our analysis is the fact that throughout this paper we shall consider both the parameters of the theory namely, $ \beta^{2} $ and $ \Theta_{2} $ are much less than unity \cite{Cadoni:2012uf} namely $ \beta^{2}\ll 1$ and $|\Theta_{2}|\ll 1 $. Therefore in principle we can always ignore terms (compared to that with the unity) those are essentially at most at the quadratic or in any subsequent higher orders in these parameters.} which obey certain constraint among themselves namely, $ \Theta_{1}=\frac{1}{\Theta_{2}}-\Theta_{2} $.

Eq.(\ref{E3}) essentially is the starting point of our analysis. In order to start our explorations on the hydrodynamic description of the above black brane configuration (\ref{E3}), it is customary to discuss some of the essential thermodynamic properties which will be required in the subsequent analysis. The Hawking temperature as well as the thermal entropy (density) of the SBB solution (\ref{E3}) turns out to be \cite{Cadoni:2012uf},
\begin{eqnarray}
T = \frac{3\gamma \Theta_{2}^{\frac{3\beta^{2}-1}{3 \gamma}}(1-\Theta_{2}^{2})^{1/3}}{4 \pi(1+3\beta^{2})};~~s=4 \pi \Theta_{2}^{-2/3\gamma}(1-\Theta_{2}^{2})^{2/3}.
\end{eqnarray}

Before we actually conclude this Section, for the sake of completeness, a few important remarks and/or comments are in order and these are essentially related to the string theory realization of the SBB solutions (\ref{E2}) in general in any dimension ($ d $). In \cite{Cadoni:2012ea}, it has been argued that SBB solutions (\ref{E2}) could in principle be obtained from black $ p $- branes by means of Kaluza-Klein (KK) compactification on a sphere ($ S^{q},~~q=d-p-2 $). 

Let us illustrate this point a bit further. Usually in $ d $ dimensions one starts with the bosonic action of the form,
\begin{eqnarray}
I = \int d^{d}x \left( R - \frac{1}{2}(\partial \Phi)^{2}-\frac{e^{a \Phi}}{2(p+2)!}F_{(p+2)}\right) 
\end{eqnarray}
where $ F_{(p+2)} $ is the $ (p+2) $ form and $ \Phi $ is the dilaton. The black $ p $- branes could be expressed as the Ramond-Ramond charged solution of the above SUGRA theory which could be formally expressed as \footnote{For details see \cite{Cadoni:2012ea}.},
\begin{equation}
ds^{2}=H(r)^{\frac{-2 \tilde{D}}{\rho}}\left(h(r)dt^{2}+\sum_{i=1}^{p}dx_{i}^{2} \right)+ H(r)^{\frac{2 D}{\rho}}(h^{-1}dr^{2}+r^{2}d\Omega_{q}^{2}).
\end{equation}

Finally, using the following KK compactification,
\begin{eqnarray}
ds^{2}= e^{-\frac{2q}{p}\psi}ds^{2}_{p+2}+e^{2\psi} d\Omega_{q}^{2}
\end{eqnarray}
and doing some trivial manipulation one can in fact express the metric $ ds^{2}_{p+2} $ in the above form (\ref{E2}) with some proper identification of the parameter $ \beta^{2} $.

\section{$ R $- Charge diffusion}
Using the methods of the $AdS_{4}/CFT_3$ duality, in this section our goal is to compute the DC electrical conductivity ($ \sigma_{DC} $) for certain special class of QFTs that come up with an additional UV length scale along with the $ z=1 $ critical exponent and the negative hyperscaling violating parameter ($ \theta <0 $). As mentioned earlier in the introduction, near the extremality, the IR physics of these QFTs could be described in terms of infra-red fixed point and which is therefore considered to be scale invariant with $ z=1 $ and $ \theta =0 $. These QFTs are quite unusual both from the point of view of the earlier holographic models \cite{Dong:2012se},\cite{Kim:2012nb},\cite{Huijse:2011ef} as well as from the field theories encountered in the usual condensed matter systems \cite{Fisher:1986zz}. The reason for this is that there one expects the hyperscaling violation to occur in the IR instead of the UV. Since these are the systems where the perturbative calculations do not seem to work very well, therefore the holographic realization of the charge transport phenomena would be indeed an interesting subject in itself.

 The entity that we are finally interested in is to compute the $ R $-charge diffusion ($ \mathcal{D} $) using the so called \textit{Einstein's relation} namely, $ \mathcal{D}= \sigma_{DC}/\chi$ (in the so called \textit{hydrodynamic limit}) for these special class of strongly coupled QFTs considering the SBBs (\ref{E3}) in the dual gravitational counterpart. In order to compute the DC conductivity, one essentially needs to evaluate the retarded two point current-current correlation \footnote{Note that here $ e $ is the small gauge coupling parameter which has been introduced by considering the fact that the corresponding $ U(1) $ symmetry at the boundary is \textit{weakly} gauged \cite{Kovtun:2008kx}.},
\begin{eqnarray}
\sigma_{DC} &=& - e^{2}\lim_{\omega \rightarrow 0}\frac{1}{\omega}Im\ \mathcal{G}^{R}_{x,x}(\omega , \textbf{q}=0)\nonumber\\
\mathcal{G}^{R}_{x,x}(\omega , \textbf{q}=0) &=& -i \int d\tau\ d\textbf{x}\ e^{i\omega \tau}\ \Theta (t)\ \langle[J_x (\textbf{x}), J_x (0)]\rangle \label{E5}
\end{eqnarray}
over the fixed background (\ref{E3}). One can perform this computation following the original prescription of \cite{Kovtun:2003wp}-\cite{Policastro:2002se} where one usually treats the Maxwell action,
\begin{eqnarray}
S_M = -\frac{1}{4g^{2}_{SG}}\int d^{4}x \sqrt{-g}\mathcal{F}_{ab}\mathcal{F}^{ab}\label{E6}
\end{eqnarray}
as perturbations over the non extremal background (\ref{E3}).

The first step towards computing the $ R $- charge diffusion is to compute the charge susceptibility ($ \chi (=\varrho/\mu) $) for which we need to know both the charge density ($ \varrho $) as well as the chemical potential ($ \mu $) for the boundary theory. Following the $AdS_4/CFT_3$ prescription \cite{Kovtun:2008kx}, the charge density for the present theory turns out to be,
\begin{eqnarray}
\varrho &=&\lim_{r\rightarrow \infty} \frac{\delta S^{(OS)}_{M}}{\delta \mathcal{A}_{t}}=\frac{e^{2}\mu \mathcal{Z}(1+3\beta^{2})}{3\gamma^{3} g^{2}_{SG}L^{2}}\nonumber\\
\mathcal{Z}&=&\frac{\mathcal{C} \Theta _2 \left(\left(\frac{1}{\Theta _2}-\Theta _2\right){}^{1/\alpha }\right){}^{\frac{2 \alpha  \beta ^2}{3 \gamma }+\frac{2}{3 \beta ^2+1}-1} \left(\left(\left(\frac{1}{\Theta _2}-\Theta _2\right){}^{1/\alpha }\right){}^{\alpha }+\Theta _2\right){}^{-\frac{2 \beta ^2}{3 \gamma }-1}}{\, _2F_1\left(1,\frac{3 (\alpha +1) \beta ^2+\alpha -1}{3 \alpha  \beta ^2+\alpha };\frac{3 \beta ^2-1}{3 \alpha  \beta ^2+\alpha }-\frac{2 \beta ^2}{3 \gamma }+1;-\frac{\left(\left(\frac{1}{\Theta _2}-\Theta _2\right){}^{1/\alpha }\right){}^{\alpha }}{\Theta _2}\right)}\nonumber\\
\mathcal{C}&=&\beta ^2 \left(\alpha  \left(6 \beta ^2+2\right)-9 \gamma \right)+3 \gamma . \label{E7}
\end{eqnarray}

Using (\ref{E7}), one can immediately read off the charge susceptibility as,
\begin{eqnarray}
\chi = \frac{e^{2} \mathcal{Z}(1+3\beta^{2})}{3\gamma^{3} g^{2}_{SG}L^{2}}.\label{E8}
\end{eqnarray}

Finally, we are in a position to compute the DC conductivity which could be performed by turning on fluctuations in the spatial component(s) of the $ U(1) $ gauge field namely,
\begin{eqnarray}
\mathcal{A}_{x}(r,t) \sim \int d\omega e^{-i \omega t} \mathcal{A}_{x}(r)
\end{eqnarray}
where $ \mathcal{A}_{x}(r) $  satisfies the equation of the following form namely,
\begin{eqnarray}
\mathcal{A}''_{x}+\frac{b'(r)}{b(r)}\mathcal{A}'_{x}-\omega^{2}\frac{c(r)}{b(r)}\mathcal{A}_{x}=0 \label{E10}
\end{eqnarray}
where, $ b(r)=\sqrt{-g}g^{rr}g^{xx} $ and $ c(r)=\sqrt{-g}g^{xx}g^{tt} $.

Considering the ingoing wave boundary condition \cite{Policastro:2002se} near the horizon of the SBBs (\ref{E3}), the natural next step would be to solve the above equation (\ref{E10}) in the low frequency ($ |\omega| \ll 1 $) regime. In order to do that we consider the following ansatz namely,
\begin{eqnarray}
\mathcal{A}_{x}=(1-(r_H/r)^{\alpha})^{\lambda}\mathcal{X}(r)\label{E11}
\end{eqnarray}
where $ r_H/r $ is a dimensionless entity with $ r_H = \left(\frac{1}{\Theta _2}-\Theta _2\right){}^{1/\alpha } $ as the position of the horizon \cite{Cadoni:2012uf}. The parameter $ \lambda $ could be estimated by imposing the so called incoming wave boundary condition near the horizon of the SBBs (\ref{E3}) which finally yields,
\begin{eqnarray}
\lambda &=& - i \mathcal{N}\omega \nonumber\\
 \mathcal{N} &=& \frac{r_{H}^{ \alpha -\frac{2}{3 \beta ^2+1}}}{\alpha}\left(\frac{1- 2\Theta_{2}^{2}}{1- \Theta_{2}^{2}} \right)^{\frac{2\beta^{2}}{3 \gamma}}.\label{E12}
\end{eqnarray}

In order to solve (\ref{E10}), we consider the following perturbative expansion of the function $\mathcal{X}(r) $ namely,
\begin{eqnarray}
\mathcal{X}(r)=\mathcal{X}^{(0)}(r)+ i \omega \mathcal{X}^{(1)}(r)+\mathcal{O}(\omega^{2}).\label{E13}
\end{eqnarray}
Substituting (\ref{E11}) into (\ref{E10}) together with (\ref{E12}) and (\ref{E13}) we arrive at the following set of equations namely,
\begin{eqnarray}
\mathcal{X}^{''(1)}-\frac{2 \mathcal{N}\alpha r_H^{\alpha}}{r^{\alpha+1}(1-r^{\alpha}_{H}/r^{\alpha})}\mathcal{X}^{'(0)}+\frac{b'}{b}\mathcal{X}^{'(1)}\nonumber\\
+\frac{\mathcal{N}\alpha r_H^{\alpha}}{r^{\alpha+1}(1-r^{\alpha}_{H}/r^{\alpha})}\left( \frac{\alpha +1}{r}+\frac{\alpha r_H^{\alpha}}{r^{\alpha+1}(1-r^{\alpha}_{H}/r^{\alpha})}-\frac{b'}{b}\right)\mathcal{X}^{(0)}&=&0\nonumber\\
\mathcal{X}^{''(0)}+\frac{b'}{b}\mathcal{X}^{'(0)}&=&0. \label{E14}
\end{eqnarray}

The natural next task would be to solve the above set of equations (\ref{E14}). As the exact analytic solutions to the above set of equations (\ref{E14}) turn out to be quite difficult to achieve, therefore in the following we only note down the corresponding solutions in the large radius limit ($ r_H <<r $) namely\footnote{Note that in the large radius limit we retain only linear order terms in the ratio $ r_H^{\alpha}/r^{\alpha +1} $.},
\begin{eqnarray}
\mathcal{X}^{(0)}(r)&=&\frac{\mathcal{C}_1 \left(3 \beta ^2 r+r\right)^{1-\frac{2}{3 \beta ^2+1}}}{3 \beta ^2-1}+\mathcal{C}_2\nonumber\\
\mathcal{X}^{(1)}(r) &=& \mathcal{C}_4+\frac{r^{-\alpha } \left(3 \beta ^2 r+r\right)^{-\frac{2}{3 \beta ^2+1}} \left(\left(\frac{1}{\Theta _2}-\Theta _2\right){}^{1/\alpha }\right){}^{-\frac{2}{3 \beta ^2+1}}}{\alpha  \left(3 \beta ^2-1\right)} \mathcal{F}(r)
\label{E15}
\end{eqnarray} 
where the function $ \mathcal{F}(r) $ could be formally expressed as,
\begin{eqnarray}
\mathcal{F}(r)=\alpha  \left(3 \beta ^2+1\right) \mathcal{C}_3 r^{\alpha +1} \left(\left(\frac{1}{\Theta _2}-\Theta _2\right){}^{1/\alpha }\right){}^{\frac{2}{3 \beta ^2+1}}
\nonumber\\
-\left(3 \beta ^2+1\right) \mathcal{C}_1 r \left(\left(\frac{1}{\Theta _2}-\Theta _2\right){}^{1/\alpha }\right){}^{2 \alpha } \left(\frac{2 \Theta _2^2-1}{\Theta _2^2-1}\right){}^{\frac{2 \beta ^2}{3 \gamma }}
\nonumber\\
-\left(3 \beta ^2-1\right) \mathcal{C}_2 \left(\left(\frac{1}{\Theta _2}-\Theta _2\right){}^{1/\alpha }\right){}^{2 \alpha } \left(3 \beta ^2 r+r\right)^{\frac{2}{3 \beta ^2+1}} \left(\frac{2 \Theta _2^2-1}{\Theta _2^2-1}\right){}^{\frac{2 \beta ^2}{3 \gamma }}.
\end{eqnarray}

Using (\ref{E15}), the DC conductivity (\ref{E5}) finally turns out to be,
\begin{eqnarray}
\sigma_{DC} = \frac{e^{2}}{2 g^{2}_{SG}}\frac{(\mathcal{C}_{2}\mathcal{C}_{3}-\mathcal{C}_{1}\mathcal{C}_{4})}{(1+3\beta^{2})^{\frac{2}{1+3\beta^{2}}}}.\label{E17}
\end{eqnarray}

Using (\ref{E8}) and (\ref{E17}) it is in fact quite trivial to compute the $ R $- charge diffusion constant for the boundary hydrodynamics which turns out to be,
\begin{eqnarray}
\mathcal{D}=\frac{3\gamma^{3}L^{2}(\mathcal{C}_{2}\mathcal{C}_{3}-\mathcal{C}_{1}\mathcal{C}_{4})}{2 \mathcal{Z}(1+3\beta^{2})^{\frac{3(1+\beta^{2})}{1+3\beta^{2}}}} = \mathfrak{K}~ T^{\frac{\theta -3}{\theta- 1}}
\label{E18}
\end{eqnarray}
where, $ \theta =\frac{6\beta^{2}}{3\beta^{2}-1} $ stands for the hyperscaling violating exponent for the boundary QFT \cite{Cadoni:2011yj}-\cite{Cadoni:2012ea} and $\mathfrak{K} $ corresponds to the overall numerical factor and/or the $ c $- number sitting in front of the temperature. At this stage it is in fact quite interesting to note that since the spatial component of the gauge field has the inverse dimension of length namely, $ [\mathcal{A}_x] =L^{-1}$ therefore the arbitrary coefficients $ \mathcal{C}_{i} (i=1,2,3,4) $ also carry the same dimension namely, $ [\mathcal{C}_{i}]=L^{-1} $ and as a result of this, the $ R $- charge diffusion (\ref{E18}) is effectively a dimensionless entity and/or a pure $ c $- number near the UV scale of the boundary QFT in the strong coupling regime. Finally, it is worthwhile to point out that unlike the case for the AdS black branes \cite{Kovtun:2003wp}-\cite{Policastro:2002se} (i.e; space times with AdS asymptotics), the charge diffusion constant (\ref{E18}) goes under a different scaling with the temperature and this might be regarded as the consequence of the broken scale invariance near the UV scale of the theory. This suggest that the QFTs that we consider in this paper should fall under a separate universality class.
 
\section{$ \eta/s $}
Having done the computation on the $ R $- charge diffusion, in this section we turn our attention towards computing the two point correlation between the stress energy tensor of the boundary QFT in the hydrodynamic limit. Looking at this issue from a broader perspective, our goal is to make a systematic comparison between the $ \eta/s $ ratios corresponding to these scalar dressed black brane configurations considering both the extremal as well as the non extremal limits. We shall concentrate on the non extremal case first.  

 According to the $ AdS_4/CFT_3 $ prescription, in order to compute the two point correlation between the boundary stress tensor what one essentially needs to do is to turn on the metric fluctuations ($ h_{xy} $) in the bulk since the boundary value of the graviton fluctuations ($ h^{(0)}_{xy} $) essentially acts as the source for the energy momentum tensor of the boundary QFT. In Quantum Field Theories although there are in principle several ways to compute the retarded correlators (namely, the Green's functions), however in our analysis we shall adopt the so called Kubo's formula that stands for the most elegant as well as the straight forward way to compute the two point function between the stress tensor that finally yields the so called hydrodynamic transport/response parameter namely the coefficient of viscosity \footnote{At this stage it is customary to note that the metric perturbations with one spatial index namely $ y $ are essentially the vectors with respect to the boundary global $ SO(2) $ symmetry where as on the other hand modes with two or no $ y $ indices are the scalars provided we had turned on our spatial momentum along $ x $ direction.}\cite{Policastro:2001yc}-\cite{Iqbal:2008by},
\begin{eqnarray}
\eta = -\lim_{\omega \rightarrow 0} \frac{1}{\omega} Im\ \mathcal{G}^{R}_{xy,xy}(\omega, \textbf{q}=0).
\end{eqnarray} 

The first step towards computing the retarded correlator is to turn on the metric fluctuations of the following form namely \footnote{At this stage one could easily check that the graviton fluctuations ($ h_{xy} $) do not mix up with other perturbations of the theory, for example the scalar perturbations ($ \delta \phi $). Therefore at least for shear mode calculations these fluctuations are not important. In fact the scalar fluctuations mix up non trivially with the graviton perturbations in the bulk while studying the sound propagation in a fluid medium and therefore it plays an important role there. Finally, it is noteworthy to mention that the effective action corresponding to graviton fluctuations turns out to be at most second order in derivatives of $ \Phi (r,t)$.},
\begin{eqnarray}
h_{xy}(r,t)&=& \Delta^{\frac{2\beta^{2}}{3\gamma}}r^{\frac{2}{1+3 \beta^{2}}} \Phi (r,t)\nonumber\\
\Phi (r,t)&=& \int d\omega e^{-i \omega t}\Phi_{\omega}(r).\label{E20}
\end{eqnarray}

As a natural next step we substitute (\ref{E20}) into the linearised Einstein's equation,
\begin{eqnarray}
\mathcal{R}^{(1)}_{ab}=\frac{1}{2}h_{ab}V(\phi)\label{E21}
\end{eqnarray}
which finally yields the differential equation of the form,
\begin{eqnarray}
-\frac{\left(3 \beta ^2+1\right)^2 \Theta _2^2 \omega ^2 r^{\frac{3}{3 \beta ^2+1}} \Phi_{\omega} (r)}{\Theta _2 r^{\frac{3}{3 \beta ^2+1}}+\left(\Theta _2^2-1\right) r^{\frac{3 \beta ^2}{3 \beta ^2+1}}}+\frac{r^{-\frac{2}{3 \beta ^2+1}-2} \left(\Theta _2 r^{1-\frac{4}{3 \beta ^2+1}}+1\right){}^{\frac{4 \beta ^2}{3 \left(\beta ^2-1\right)}}}{\left(\Theta _2 r^{\frac{3 \beta ^2}{3 \beta ^2+1}}+r^{\frac{3}{3 \beta ^2+1}}\right){}^2}\Bbbk (r)=\left(\beta ^2-1\right)^2 \Theta _2 L^2 h_{xy}V(\phi)\nonumber\\
\label{E22}
\end{eqnarray}
where the function $ \Bbbk (r) $ could be formally expressed as,
\begin{eqnarray}
\Bbbk (r)=\left(3 \beta ^2+1\right)^2 \Theta _2^2 \left(\Theta _2^2-1\right) \left(-r^4\right) \left(r \Phi_{\omega} ''(r)+\Phi_{\omega} '(r)\right)-6\Theta _2 r^{\frac{4}{3 \beta ^2+1}+2}\left(\beta ^2-1\right) \left(\beta ^2-\Theta _2^2\right) \Phi_{\omega} (r)\nonumber\\
-\Theta _2 r^{\frac{12}{3 \beta ^2+1}} \left(\left(3 \beta ^2+1\right) r \left(\left(3 \beta ^2+1\right) r \Phi_{\omega} ''(r)+4 \Phi_{\omega} '(r)\right)-6 \left(\beta ^2-1\right) \Phi_{\omega} (r)\right)\nonumber\\
+\Theta _2 r^{\frac{4}{3 \beta ^2+1}+2} \left(\left(3 \beta ^2+1\right) r \left(2 \left(-3 \left(\beta ^2+1\right) \Theta _2^2+3 \beta ^2+1\right) \Phi_{\omega} '(r)-\left(3 \beta ^2+1\right) \left(3 \Theta _2^2-2\right) r \Phi_{\omega} ''(r)\right)\right)\nonumber\\
+3r^{\frac{8}{3 \beta ^2+1}+1} \Theta _2^2 \left(\left(3 \beta ^2+1\right) r \left(-\left(3 \beta ^2+1\right) r \Phi_{\omega} ''(r)-\left(\beta ^2+3\right) \Phi_{\omega} '(r)\right)+4 \left(\beta ^2-1\right) \Phi_{\omega} (r)\right)\nonumber\\
+r^{\frac{8}{3 \beta ^2+1}+1}\left(3 \beta ^2+1\right)^2 r \left(r \Phi_{\omega} ''(r)+\Phi_{\omega} '(r)\right).
\end{eqnarray}

Regarding the above equation in (\ref{E22}) the two crucial points are to be noted. Firstly, since we are interested to evaluate the retarded Green's function exactly at the boundary therefore we do not actually have to bother about the potential $ V(\phi) $ in our equation as it approaches zero near the boundary ($ r_H/r \ll 1 $) \cite{Cadoni:2012uf}. Secondly, it looks exactly as that of the wave equation for a scalar minimally coupled to the curved background. In order to solve (\ref{E22}) we take the following ansatz namely,
\begin{eqnarray}
\Phi_{\omega}(r)=\left(1-r_H^{\alpha}/r^{\alpha} \right)^{\delta} \mathcal{H}(r)\label{E24} 
\end{eqnarray} 
where $ \delta $ could be fixed by the incoming wave boundary condition,
\begin{eqnarray}
\delta &=& - i \Gamma \omega\nonumber\\
 \Gamma &=& \frac{r_H^{(1-\alpha )/2}}{\alpha}\left(1+\frac{\Theta_{2}}{r_H^{\alpha}} \right)^{\frac{2\beta^{2}}{3(1-\beta^{2})}}\label{E25}
\end{eqnarray}
subjected to the following constraint satisfied at the horizon of the black brane namely,
\begin{eqnarray}
2\Theta_{2}^{3}+3\Theta_{2}^{2}r_H^{\alpha}+\Theta_{2}r_H^{2\alpha}-r_H^{\alpha +1}(2+r_H^{\alpha})=0
\end{eqnarray}
which is essentially a polynomial in $ \Theta_{2} $ that basically constraints the value  of the dimensionless parameters $ \Theta_{i}(i=1,2) $ of our theory in order to make it to be consistent with the ingoing wave boundary condition near the horizon of the black brane.

Finally, considering (\ref{E25}) and substituting (\ref{E24}) into (\ref{E22}) we arrive at the following equation namely,
\begin{eqnarray}
\mathcal{C}_{1}(r)\mathcal{H}''(r)+\mathcal{C}_2(r)\mathcal{H}'(r)+\mathcal{C}_{3}(r)\mathcal{H}(r)=0 \label{E27}
\end{eqnarray}
where each of the individual coefficients $ \mathcal{C}_i (r)$($ i=1,2,3 $) could be read off as,
\begin{eqnarray}
\mathcal{C}_1=-(1+3\beta^{2})\Theta_{2}r^{5+\alpha}(r^{\alpha}+\Theta_{2})^{2}\nonumber
\end{eqnarray}
\begin{eqnarray}
\mathcal{C}_2 = 2i\Gamma \omega \alpha r_H^{\alpha}\Theta_{2}(1+3\beta^{2})(r^{\alpha}+\Theta_{2})^{2} r^{4}+(1+3\beta^{2})\Theta_{2}^{3}r_H^{\alpha}r^{4}-6(1+\beta^{2})\Theta_{2}^{3}r^{\alpha +4}\nonumber\\
+ 2\Theta_{2}(1+3\beta^{2})r^{\alpha +4}-4\Theta_{2}r^{4+3\alpha}-3\Theta_{2}^{2}(\beta^{2}+3)r^{2\alpha +4}+(1+3\beta^{2})r^{2\alpha +4}\nonumber
\end{eqnarray}
\begin{eqnarray}
\mathcal{C}_3 = -i\Gamma \omega \alpha r_H^{\alpha}\left[ (\alpha +1)(1+3\beta^{2})\Theta_{2}(r^{\alpha}+\Theta_{2})^{2}r^{3}+(1+3\beta^{2})\Theta_{2}^{3}r^{3}r_H^{\alpha}-6\Theta_{2}^{3}(1+\beta^{2})r^{3}\right]\nonumber\\
 -i\Gamma \omega \alpha r_H^{\alpha}\left[ 2\Theta_{2}(1+3\beta^{2})r^{3}-4\Theta_{2}r^{2\alpha +3}-3\Theta_{2}^{2}(\beta^{2}+3)r^{\alpha+3}+(1+3\beta^{2})r^{\alpha +3}\right]\nonumber\\
 -6\Theta _2 r^{3+\alpha}\frac{\left(\beta ^2-1\right)}{(1+3\beta^{2})}\left(\beta ^2-\Theta _2^2-r^{2\alpha}-2\Theta_{2}r^{\alpha}\right). 
\end{eqnarray}
Note that in the above equation (\ref{E27}) we have retained terms only upto leading order in the frequency $ \omega $ as the higher order terms do not contribute to the coefficient of viscosity ($ \eta $). Our next task would be to solve (\ref{E27}) perturbatively in the frequency $ \omega $ namely,
\begin{eqnarray}
\mathcal{H}(r) &=& \mathcal{H}^{(0)}(r)+ i \omega \mathcal{H}^{(1)}(r)+\mathcal{O}(\omega^{2})\nonumber\\
\mathcal{C}_i (r)&=& \mathcal{C}^{(0)}_i (r)+i \omega \mathcal{C}_i^{(1)} (r).\label{E29}
\end{eqnarray}

Substituting (\ref{E29}) into (\ref{E27}) we arrive at the following set of equations,
\begin{eqnarray}
\mathcal{C}_{1}^{(0)}\mathcal{H}''^{(0)}+\mathcal{C}_{2}^{(0)}\mathcal{H}'^{(0)}+\mathcal{C}_{3}^{(0)}\mathcal{H}^{(0)}&=&0\nonumber\\
\mathcal{C}_{1}^{(0)}\mathcal{H}''^{(1)}+\mathcal{C}_{2}^{(0)}\mathcal{H}'^{(1)}+\mathcal{C}_{2}^{(1)}\mathcal{H}'^{(0)}+\mathcal{C}_{3}^{(0)}\mathcal{H}^{(1)}+\mathcal{C}_{3}^{(1)}\mathcal{H}^{(0)}&=&0.\label{E30}
\end{eqnarray}

The solutions corresponding to (\ref{E30}) turn out to be\footnote{Note that as the exact solutions corresponding to (\ref{E30}) is indeed quite non trivial to achieve therefore in order to obtain these solutions (\ref{E31}) we have taken into account a few important facts, for example, we have dropped all the terms of the type $ \Theta_{2}/r^{\alpha} $ including its various higher powers in the large $ r $ limit as $ |\Theta_{2}|\ll 1 $ \cite{Cadoni:2012uf}. Moreover in our computation we have dropped all the terms $ \sim 1/r^{3} $ on wards. Finally, here $ \mathcal{K}_{i} $s are some arbitrary coefficients that could be fixed in terms of asymptotic normalization conditions namely, $ \Phi_{\omega}(\infty)=1 $.},
\begin{eqnarray}
\mathcal{H}^{(1)}  \approx  \mathcal{H}^{(0)} = \mathcal{K}_{2}- \mathcal{K}_{1}\frac{7  \Theta _2 e^{\frac{16 \Theta _2 r_H^{\alpha}-7}{9 \Theta _2 r^{\alpha}}}}{16 \Theta _2 r_H^{\alpha}-7}.\label{E31}
\end{eqnarray}

Using the above solution (\ref{E31}) as well as the asymptotic boundary condition, the coefficient of viscosity finally turns out to be,
\begin{eqnarray}
\eta = \Theta_{2}^{-2/3\gamma}(1-\Theta_{2}^{2})^{2/3}
\end{eqnarray}
which finally yields,
\begin{eqnarray}
\frac{\eta}{s} = \frac{1}{4 \pi}.\label{E33}
\end{eqnarray}

Eq.(\ref{E33}) is expected on general ground for theories coupled to (two derivative) Einstein's theory of gravity in the non extremal limit. The key feature of the present analysis is that it ensures the above universality (\ref{E33}) for space times that asymptotically approaches towards a DW kind of geometry rather than the usual $ AdS_{4} $ asymptotics. It is therefore an artefact of the earlier analysis for space times with usual $ AdS_{4} $ asymptotics, namely the $ \eta/s $ ratio in the so called non extremal limit is sensitive only to the gravitational couplings of the theory and therefore does not get corrected due to the presence of the various matter couplings (non trivial matter interactions) in the theory that do not include any explicit or implicit function of the Riemann curvature \cite{Brustein:2008cg}. 

Before we conclude this section, it is also important to understand the significance of the above result from the point of view of the boundary field theory. The SBBs (\ref{E3}) in the bulk essentially describe a completely different gauge theory at the boundary rather than a different phase in the same gauge theory. In other words, the QFTs dual to SBBs are completely different from that of the QFTs dual to the usual SAdS black branes in the bulk for $ T<T_c $ \cite{Cadoni:2012ea}. More precisely, the UV physics of these dual QFTs is associated with an additional length scale in the presence of a negative hyperscaling violating factor namely, $ z=1, \theta =\frac{6\beta^{2}}{3\beta^{2}-1} $ \cite{Cadoni:2011yj}-\cite{Cadoni:2012ea}. Therefore the Gauge/gravity duality plays an important role in order to ensure the universality of $ \eta/s $ ratio for a wide class of gauge theories that might exhibit hyperscaling violation at UV.

\section{Extremal limit}
In this section, based on the standard frame work of the $ AdS_{4}/CFT_3 $ duality, we turn our attention towards the hydrodynamic description of the extremal SBBs in ($ 3+1 $) dimensions\footnote{The present analysis of this section might be considered as an extension of the earlier calculations \cite{Edalati:2009bi}-\cite{Paulos:2009yk} to space times with \textit{non} AdS asymptotics, for example the Domain Wall (DW) kind of geometry that emerges as the dual to certain class of quantum critical systems which seem to posses hyperscaling violating phase associated with some emerging UV length scale \cite{Cadoni:2011yj}-\cite{Cadoni:2012ea}.} \cite{Cadoni:2011yj}-\cite{Cadoni:2012uf}. The extremal (zero temperature) limit of SBBs could be formally obtained by setting $ \Theta_{1}=0 $ \cite{Cadoni:2012uf},
\begin{eqnarray}
ds^{2}= \Delta^{\frac{2\beta^{2}}{3\gamma}}\left[r^{\frac{2}{1+3\beta^{2}}} (-dt^{2}+d\textbf{x}^{2})+\frac{\gamma^{2}L^{2}\Theta_{2}^{\frac{-2\beta^{2}}{\gamma}}}{(1+3\beta^{2})^{2}}r^{\frac{-2}{1+3\beta^{2}}}\Delta^{\frac{4\beta^{2}}{3\gamma}}dr^{2}\right].\label{E34} 
\end{eqnarray}

For $ \Theta_{2}>0 $, Eq.(\ref{E34}) eventually corresponds to a regular scalar soliton interpolating between a scale covariant theory at UV and an $ AdS_{4} $ geometry corresponding to the IR fixed point. In order to proceed further we define a new coordinate,
\begin{eqnarray}
u =1 + r^{\alpha};~~\beta^{2}<1/3
\end{eqnarray}
which finally helps us to reduce the above metric (\ref{E34}) into the following form namely,
\begin{eqnarray}
ds^{2}=g_{tt}^{(0)}dt^{2}+g_{uu}^{(0)}du^{2}+g_{ij}^{(0)}dx^{i}dx^{j}\label{E36}
\end{eqnarray}
where the individual zeroth order coefficients of the metric reads as,
\begin{eqnarray}
g_{tt}^{(0)}&=&-(u-1)^{\frac{2}{3}}( u-1+\Theta_{2})^{\frac{2\beta^{2}}{3\gamma}} =-(u-1)^{\frac{2}{3}} \mathcal{S}_{(t)}(u)\nonumber\\
g_{ij}^{(0)}&=&(u-1)^{\frac{2}{3}} \mathcal{S}_{(t)}(u)\delta_{ij}\nonumber\\
g_{uu}^{(0)}&=&(u-1)^{-2}\left[ \frac{L^{2}}{9}\Theta_{2}^{\frac{-2\beta^{2}}{\gamma}}(u-1+\Theta_{2})^{\frac{2\beta^{2}}{\gamma}}\right] =(u-1)^{-2} \mathcal{S}_{(u)}(u)\label{E37}
\end{eqnarray} 
such that the horizon of the black brane is located at $ u=1 $ and the boundary is located at $ u\rightarrow \infty $. At this stage one should take a note on the fact that the functions $ \mathcal{S}(u) $ are perfectly regular at the horizon. The extremal nature of the black brane could be found out through the existence of the double pole in the metric component $ g_{uu}^{(0)} $ at the horizon and this feature is identical to that with the black brane solutions with usual $ AdS_{4} $ asymptotics \cite{Edalati:2009bi}-\cite{Chakrabarti:2009ht}.

At this stage it is noteworthy to mention that in the extremal limit, the near horizon geometry of SBBs (\ref{E36}) are significantly different from those of the most general class of black holes with the usual $ AdS_4 $ asymptotics. The near horizon geometry for the most generic class of extremal AdS black branes usually appears with a double pole singularity in $ g_{uu}^{(0)} $ and a double zero in $ g_{tt}^{(0)} $ \cite{Chakrabarti:2009ht}. On the other hand, for extremal SBBs (\ref{E36}) we note that the zero of the $ g_{tt}^{(0)} $ appears with a different (fractional) power which therefore distinctly categorize the extremal SBBs from that of the usual AdS black branes in the extremal limit.   

With the above (zeroth order) metric coefficients (\ref{E37}) in hand, our next task would be to consider the graviton fluctuations of the form $ g_{\mu\nu}=g_{\mu\nu}^{(0)}+h_{\mu\nu}  $, in particular,
\begin{eqnarray}
h_{xy}(u,t)&=& (u-1)^{\frac{2}{3}}\mathcal{S}_{(t)}(u)\Phi (u,t)\nonumber\\
\Phi (u,t)&=& \int d\omega e^{-i \omega t}\Phi_{\omega}(u).\label{E38}
\end{eqnarray}

Substituting (\ref{E38}) into linearized Einstein's equation (\ref{E21}) we arrive at the following differential equation namely,
\begin{eqnarray}
\frac{(u-1)^{2/3} \Theta _2^{-\frac{2 \beta ^2}{\beta ^2-1}} \left(\Theta _2+u-1\right){}^{\frac{3-2 \beta ^2}{3 \left(\beta ^2-1\right)}}\Re (u)}{4 \left(\beta ^2-1\right)^2 }-\frac{1}{2} \omega ^2 L^2 \Phi_{\omega} (u)=\frac{L^2(u-1)^{\frac{2}{3}}}{2}\mathcal{S}_{(t)}(u)\Phi_{\omega} (u) V(\phi)\label{Eq43}
\end{eqnarray} 
where the function $ \Re (u) $ could be formally expressed as,
\begin{eqnarray}
\Re (u) = 12 \left(\beta ^2-1\right) \Phi_{\omega} (u) \left(\Theta _2+u-1\right){}^{\frac{1}{\beta ^2-1}} \left(-(\beta -1) \Theta _2+u-1\right) \left((\beta +1) \Theta _2+u-1\right)\nonumber\\
-18 \left(\beta ^2-1\right)^2 (u-1) \left(\Theta _2+u-1\right){}^{\frac{2 \beta ^2-1}{\beta ^2-1}} \left((u-1) \Phi_{\omega} ''(u)+2 \Phi_{\omega} '(u)\right).\label{E40}
\end{eqnarray}

Following the original prescription of \cite{Edalati:2009bi} our next task would be to solve the above equation (\ref{Eq43}) at two extremal limits, namely close to the horizon of the extremal black brane as well as near the boundary of the space time. Finally, we would match both of these solutions at some intermediate matching region ($ u\rightarrow 1 $) \cite{Edalati:2009bi}. We first focus on the near horizon solution. In order to proceed further we define a new variable ($ \zeta $),
\begin{eqnarray}
u= 1+\frac{\omega}{\zeta}
\end{eqnarray}
and express our near horizon solution for both $ \omega L\rightarrow 0 $ and $ \zeta L \rightarrow 0 $ in such a way so that the ratio $ \frac{\omega}{\zeta}\rightarrow 0 $. 
With the choice of this new variable ($ \zeta $) the near horizon ($ u \sim 1 $) equation for $ \Phi_{\omega} $ turns out to be \footnote{At this stage it is noteworthy to mention that the $ u $ coordinate is a suitable coordinate only in the asymptotic region ($ u\gg 1 $). On the other hand, $ \zeta $ is the appropriate variable for the near horizon expansion as it is easy to note that in the near horizon limit ($ u \sim 1 $) the $ \omega $ expansion turns out to be ill defined in the $ u $ coordinate. All these features are quite reminiscent of the usual scenario that one encounters for $ AdS_{4} $  black branes \cite{Edalati:2009bi}-\cite{Chakrabarti:2009ht}.},
\begin{eqnarray}
\Phi_{\omega}'' \approx 0
\end{eqnarray}
whose solution could be formaly expressed as,
\begin{eqnarray}
\Phi_{\omega} = q_{I} (1 + i L\zeta)=q_{I} \left(1+ \frac{i \omega L}{u-1}\right).\label{E43}
\end{eqnarray}

Eq.(\ref{E43}) represents an ingoing wave near the horizon of the extremal SBB (\ref{E36}) at least if we truncate ourselves upto leading order in $ \zeta $ considering it to be vanishingly small. Our next task would be to find the precise solution near the boundary ($ u \gg 1 $) of the space time. In this case the equation corresponding to $ \Phi_{\omega}(u) $ turns out to be,
\begin{eqnarray}
\Phi_{\omega}'' +\frac{2}{u-1} \Phi_{\omega}' \approx 0. \label{E44} 
\end{eqnarray}

The solution corresponding to (\ref{E44}) could be formally expressed as,
\begin{eqnarray}
\Phi_{\omega}(u)&=& \mathcal{Q}_{I}^{(0)}-\frac{\mathcal{Q}_{I}^{(1)}}{(u-1)} \label{E45}
\end{eqnarray}
where $ \mathcal{Q}_{I}^{(0)} $ and $ \mathcal{Q}_{I}^{(1)} $ are two arbitrary integration constants that could be related to each other by the normalized boundary condition namely $\Phi_{\omega}(\infty)=1 $. Note that in order to arrive at the above equation (\ref{E44}) we have made use of the fact that $ |\Theta_{2}| \ll 1$ and $ \beta^{2}\ll 1 $ \cite{Cadoni:2012uf} such that one could neglect terms of the type $ \Theta_{2}^{2}\beta^{2} $ compared to unity. Finally, we match the solutions (\ref{E43}) and (\ref{E45}) for $ u\rightarrow 1 $ which finally yields,
\begin{eqnarray}
\mathcal{Q}_{I}^{(0)}=q_{I},~~~\mathcal{Q}_{I}^{(1)}=-iq_{I}\omega L \label{E46}
\end{eqnarray}
 Finally, substituting (\ref{E46}) into (\ref{E45}) the final form of the graviton wave function (\ref{E45}) turns out to be,
\begin{eqnarray}
\Phi_{\omega}(u)= q_{I}\left[ 1 +i  \frac{\omega L}{(u-1)}\right] .\label{E47} 
\end{eqnarray} 

Using (\ref{E47}) the retarded two point correlator finally turns out to be,
\begin{eqnarray}
\mathcal{G}^{R}_{xy,xy}= -\frac{i \omega}{16 \pi G_N}.\label{E52}
\end{eqnarray}

Using (\ref{E52}) the viscosity to entropy ratio finally turns out to be,
\begin{eqnarray}
\frac{\eta}{s}&=&\frac{1}{4 \pi}\label{53}
\end{eqnarray}
which therefore satisfies the conjectured lower bound namely, $ \eta/s \geqslant 1/4 \pi$ like in the non extremal case. From the bulk point of view this result is a bit surprising due to the fact that the time component of the metric exhibits a different (near horizon) zero limit as compared to those of the extremal (charged) AdS black branes \cite{Edalati:2009bi}-\cite{Chakrabarti:2009ht}. As an example, for $ AdS_4 $ Reissner Nordstrom black holes $ g_{tt}^{(0)} $ exhibits a double zero near the horizon of the black brane and because of this fact the near horizon geometry turns out to be $ AdS_{2} \times R^{2} $ \cite{Edalati:2009bi}-\cite{Chakrabarti:2009ht}. On the other hand, in the extremal limit of SBBs, the time component of the metric namely, $ g_{tt}^{(0)} $ exhibits a different zero limit (see (\ref{E37})) and which does not appear as the inverse of the double pole structure in $ g_{uu}^{(0)} $. As a result the IR geometry turns out to be different than the usual $ AdS_{2} \times R^{2} $.  
From the point of view of the boundary field theory this result is even more illuminating. It simply translates the fact that the quantum critical systems even in the presence of negative hyperscaling violation (near the UV scale of the theory) exhibit identical shear as of those with no hyperscaling violation ($ \theta =0 $)\cite{Edalati:2009bi}-\cite{Chakrabarti:2009ht}, or hyperscaling violation near the IR scale of the theory \cite{Dong:2012se},\cite{Kim:2012nb},\cite{Huijse:2011ef}.
 
\section{Summary and final remarks}
We now summarize the key findings of our analysis. The basic motivation behind the present analysis was to explore the hydrodynamic regime of a very special class of strongly coupled QFTs that exhibit the so called hyperscaling violation at the UV scale of the theory in the presence of a \textit{negative} hyperscaling violating exponent ($ \theta <0 $). The dual gravitational counterpart for these special class of QFTs encodes scalar dressed black branes with non AdS asymptotics \cite{Cadoni:2011yj}-\cite{Cadoni:2012ea}. In the extremal limit, the dual description in the bulk takes the form of the scalar dressed extremal soliton endowed in between a fixed point at IR and the Domain Wall (DW) geometry at UV. 

The entire analysis of this paper could in principle be divided into two parts. In the first part of our analysis we compute the $ R $- charge diffusion for these special class of strongly coupled hyperscaling violating QFTs near its UV scale. From our analysis we note that the charge diffusion coefficient scales non trivially with temperature as compared to the usual case of AdS black branes \cite{Kovtun:2003wp}-\cite{Policastro:2002se}. We identify this as the natural consequence of the hyperscaling violation ($ \theta<0 $) near the UV scale of the theory. 

In the second part of our analysis we particularly focus on the universality of $ \eta/s $ ratio both in the non extremal as well as the extremal limit. In the non extremal limit the $ \eta/s $ ratio turns out to be $ 1/4\pi $ and this essentially translates the fact that the usual membrane paradigm arguments \cite{Iqbal:2008by} also work for space times with non AdS asymptotics.

In the extremal limit the scenario turns out to be a bit surprising where we note that the conjectured lower bound for $ \eta/s $ is still satisfied. From the bulk point of view this effect should come as a surprise by noting the fact the extremal scalar soliton does not carry the explicit factor of $ AdS_2 $ in its near horizon limit and therefore the near horizon geometry is radically different from that of the extremal charged AdS black branes \cite{Edalati:2009bi}-\cite{Chakrabarti:2009ht}. However we note that the radial component of the metric ($ g_{uu}^{(0)} $) exhibits a double pole near the horizon like in the case for extremal $ AdS_{4} $ black branes. Therefore we interpret our finding (\ref{53}) as the consequence of the existence of such a pole structure near the horizon which somehow confirms the fact that membrane paradigm like arguments should always hold in the presence of a double pole singularity in the near extremal limit \cite{Chakrabarti:2009ht}. From the point of view of the boundary field this clearly explains the fact that these special class of strongly coupled quantum critical systems (with $ \theta <0 $) exhibit identical shear as those of the QFTs dual to extremal AdS black branes.

In summary, the take home message from the present analysis is that the proposed universal bound on the $ \sigma_{DC}/\chi $ ratio \cite{Kovtun:2008kx} strictly holds in the presence of the conformal invariance and trivially breaks down due to the presence of an emerging (UV) length scale. On the other hand, the $ \eta/s $ ratio is not at all sensitive to the UV sector of the theory and is entirely determined by the physics at IR and which is therefore more sacred. The final conclusion that we reach considering all these facts is that the class of gauge theories in the presence of negative hyper scaling violation ($ \theta<0 $) near the UV scale of the theory, might fall under a separate universality class as compared to those with $ \theta\geqslant 0 $. 

Before we conclude this paper a few comments are in order. It will be really a nice exercise to compute the sound modes as well as the bulk viscosity to entropy ratio for these hyperscaling violating theories employing the techniques of Gauge/gravity duality. Besides these issues, one can also repeat the above calculations for electrically charged SBBs where due to the presence of the minimal coupling with the $ U(1) $ gauge field in the theory, the near horizon structure in the extremal limit might contain an explicit $ AdS_2 $ factor and therefore this should reproduce the usual results  for the $ \eta/s $ ratio \cite{Edalati:2009bi}-\cite{Chakrabarti:2009ht}. One can also repeat the above calculations in the presence of non minimal coupling to the abelian gauge sector where one might expect the result in the extremal limit to be sensitive to the coupling function itself.

{\bf {Acknowledgements :}}
It is a great pleasure to thank Aninda Sinha for his valuable comments on the draft. The author would also like to acknowledge the financial support from CHEP, Indian Institute of Science, Bangalore.\\


\end{document}